\documentclass[aps,prd,twocolumn,amsmath,amssymb,showpacs,showkeys,floatfix]{revtex4}
\usepackage{graphicx}% Include figure files
\usepackage{dcolumn}% Align table columns on decimal point
\usepackage{bm}% bold math 
\usepackage{epsfig}
\begin{document}

\title{The effective continuum threshold in dispersive sum rules}

\author{Wolfgang Lucha$^{a}$, Dmitri Melikhov$^{a,b,c}$, and Silvano Simula$^{d}$} 

\affiliation{
$^a$Institute for High Energy Physics,
Austrian Academy of Sciences, Nikolsdorfergasse 18, A-1050, Vienna, Austria\\
$^b$Faculty of Physics, University of Vienna, Boltzmanngasse 5, A-1090, Vienna, Austria\\
$^c$D.V.~Skobeltsyn Institute of Nuclear Physics, Moscow State University, 119991, Moscow, Russia\\
$^d$INFN, Sezione di Roma Tre, Via della Vasca Navale 84, I-00146, Roma, Italy}
\begin{abstract}
We study the accuracy of the bound-state parameters obtained with the method of dispersive sum rules, 
one of the most popular theoretical approaches in nonperturbative QCD and hadron physics. 
We make use of a quantum-mechanical potential model since it provides the only possibility to probe 
the reliability and the accuracy of this method: one obtains the bound-state parameters from sum rules 
and compares these results with the exact values calculated from the Schr\"odinger equation.  
We investigate various possibilities to fix the crucial ingredient of the method of sum rules 
--- the effective continuum threshold --- and propose modifications which lead  
to a remarkable improvement of the accuracy of the extracted ground-state parameters 
compared to the standard procedures adopted in the method. Although the rigorous control of 
systematic uncertainties in the method of sum rules remains unfeasible, the application 
of the proposed procedures in QCD promises a considerable increase of the actual accuracy 
of the extracted hadron parameters. 
\end{abstract}

\maketitle
\section{Introduction}
The method of dispersive sum rules for the extraction of ground-state parameters in QCD was 
formulated in \cite{svz,ioffe} and since then has been extensively applied to the analysis 
of hadron properties \cite{footnote}.

A sum-rule calculation of hadron parameters \cite{svz,ioffe} involves two steps: 
(i) one calculates the operator product expansion (OPE) for a relevant correlator 
and formulates the sum rule which relates this OPE to the sum over hadronic states, 
and 
(ii) one attempts to extract ground-state parameters by a numerical procedure. 
Each of these steps leads to uncertainties in the final result.

The first step lies fully within QCD and allows a rigorous treatment of the 
uncertainties: the correlator in QCD is not known precisely (because of 
uncertainties in quark masses, condensates, $\alpha_s$, radiative corrections, 
etc.) but the corresponding errors in the correlator may be controlled, at least 
in principle. We refer to such errors as the {\em OPE uncertainties}.

The second step is more cumbersome: even if several terms of 
the OPE for the correlator were known precisely, the numerical procedures of sum rules 
should provide the range of values which contains the true value of the hadron parameter. 
We call this range the {\em intrinsic sum-rule uncertainty}.

In spite of the extensive applications of sum-rules in particle physics, 
including also flavor physics \cite{cern}, where a rigorous error analysis is 
mandatory, a proper investigation of the systematic uncertainties of the method 
has been started only recently \cite{lms_2ptsr,lms_3ptsr,m_lcsr}. 

The method of sum rules contains a set of prescriptions which are believed to allow 
the control of the accuracy of the extracted bound-state parameters (see 
e.g.~Ref.~\cite{standard}). 
The outcome of these prescriptions is claimed to be the estimate of the 
{\em intrinsic sum-rule uncertainty}. 

Obviously, the only possibility to acquire an unbiased judgement of the reliability of the error
estimates in sum rules is to apply the method to a problem where the parameters 
of the theory may be fixed and the corresponding parameters of the ground state 
may be calculated independently and exactly. 

Presently, only quantum-mechanical potential models provide such a possibility. 
A simple harmonic-oscillator (HO) potential model, used as a testing ground in \cite{lms_2ptsr,lms_3ptsr,m_lcsr}, 
possesses the essential features of QCD --- confinement and asymptotic freedom \cite{nsvz} --- and has 
the following advantages: 
(i) the bound-state parameters (masses, wave functions, form factors) are known precisely;
(ii) direct analogues of the QCD correlators may be calculated exactly. 

Applying the standard sum-rule machinery, we have determined the ground-state 
decay constant \cite{lms_2ptsr} and the form factor \cite{lms_3ptsr} from the relevant correlators, 
and confronted the obtained results with the known exact values, probing in this way the accuracy of the  method.
We have clearly demonstrated that the standard procedures adopted in the method of sum rules 
do not yield realistic error estimates for the extracted ground-state parameters. 
Moreover, we have shown that the uncontrolled systematic errors of the form factors are typically much 
larger than those for the decay constants. 

The natural questions which then arise are: (i) Can the ``standard'' procedures of the method of sum 
rules be modified, 
leading to an improvement of the extracted ground-state parameters? 
(ii) Can one formulate a procedure which would provide the interval surely containing the actual bound-state 
parameter? This would mean a rigorous control of the {\em intrinsic sum-rule uncertainty}. 

In this Letter, we will show that the answer to the first question is ``yes'', whereas the answer to the 
second question is ``no''. 

The crucial ingredient of sum rules is the effective continuum threshold $z_c$, which 
governs the accuracy of the quark-hadron duality hypothesis, the basic concept of the method. 
We study possible modifications of the standard procedure of fixing $z_c$. 
In the HO model, relaxing the standard assumption of a Borel-parameter independent $z_c$ is shown to lead to a 
significant improvement of the extraction of the bound-state parameters, particularly, of the form factor. 
Even though the rigorous control over the systematic uncertainties of 
the ground-state parameters obtained from sum rules is not feasible 
(and cannot be obtained in principle in problems where the truncated OPE is the only input), 
the application of our findings in QCD promises a considerable improvement of the actual accuracy of the method. 

%\vspace{-.5cm}
\section{Harmonic-oscillator model}
We consider a non-relativistic HO model defined by the Hamiltonian ($r\equiv|\vec r\,|$)
 \begin{eqnarray}
    H = H_0 + V(r) ~ , \, H_0 = {\vec p}^{\,2}/2m ~ , \, 
    V(r) = {m \omega^2 r^2}/2 ~ ,
 \end{eqnarray}
where all features of the bound states are calculable. 
For instance, for the ground (g) state one finds 
 \begin{eqnarray}
    \label{EG} 
    E_{\rm g} & = & \frac{3}{2} \omega ~ ,\quad R_{\rm g}\equiv |\Psi_{\rm g}(\vec r = 0)|^2 = 
    \left( {m \omega}/{\pi} \right)^{3/2} ~ , \nonumber\\
    \label{FG}
    F_{\rm g}(q) & = &\exp(-q^2 / 4m \omega) ~ ,
 \end{eqnarray}
where the elastic form factor of the ground state is defined according to ($q\equiv 
|\vec q|$)
 \begin{eqnarray}
    F_{\rm g}(q) = \langle \Psi_{\rm g}|J(\vec q)|\Psi_{\rm g} \rangle = \int d^3k\,
    \psi_{\rm g}^\dagger(\vec k) ~ \psi_{\rm g}(\vec k - \vec q) ~ , 
 \end{eqnarray}
with the current operator $J(\vec q)$ given by the kernel
 \begin{eqnarray}
    \label{J} 
    \langle \vec r\,'|J(\vec q)|\vec r\rangle = \exp(i\vec q \cdot \vec r) ~ 
    \delta^{(3)}(\vec r - \vec r\,') ~ .
 \end{eqnarray}

\section{Polarization operator}

In the method of dispersive sum rules the basic quantity needed for the extraction 
of the decay constant (i.e., of the ground-state wave function at the origin) is 
the correlator of two currents \cite{svz}. 
Its quantum-mechanical analogue is %\cite{nsvz}
 \begin{eqnarray}
    \label{pi} 
    \Pi(T) = \langle \vec r_f = 0 | e^{- H T} | \vec r_i = 0 \rangle ~ , 
 \end{eqnarray}
where $T$ is the Euclidean time.
In the case of the HO potential the correlator $\Pi(T)$ is exactly known: %\cite{nsvz}:
 \begin{eqnarray}
    \label{piexact} 
   \Pi(T) & = & \left( \frac{m \omega}{2\pi \mbox{sinh}(\omega T)} \right)^{3/2} ~ ,\\
   \Pi_0(T) & = & \left(\frac{m}{2\pi T}\right)^{3/2} ~ , \nonumber \\[2mm]
   \Pi_{\rm power}(T) &\equiv& \Pi(T) - \Pi_0(T) \nonumber \\ 
   & = & \left(\frac{m}{2\pi T}\right)^{3/2} \left[-\frac{1}{4}{\omega^2 T^2} + 
    \cdots \right] \nonumber ~ .
 \end{eqnarray}

\section{Vertex function}

The basic quantity for the extraction of the form factor in the method of dispersive 
sum rules is the correlator of three currents \cite{ioffe}. 
The analogue of this quantity in quantum mechanics is \cite{lms_3ptsr}
 \begin{eqnarray}
    \Gamma(\tau_2, \tau_1, q) = \langle \vec r_f = 0 | e^{-H \tau_2} J(\vec q) 
    e^{-H \tau_1} | \vec r_i = 0 \rangle ~ , 
 \end{eqnarray}
with the operator $J(\vec q)$ being defined in (\ref{J}). 
In the HO model the exact analytic expression for $\Gamma(\tau_2, \tau_1, q)$ was 
obtained in Ref.~\cite{lms_3ptsr}. 
At equal times $\tau_1 = \tau_2 = \frac12 T$ it takes the following form: 
 \begin{eqnarray}
    \label{gammaope} 
    \Gamma(T, q) & = & \Pi(T) ~ \exp\left( -\frac{q^2}{4m \omega} 
    \mbox{tanh}\left(\frac{\omega T}{2}\right) \right) ~ , \\
    \Gamma_0(T, q) & = & \Pi_0(T) ~ \exp\left( -\frac{q^2 T}{8m} \right) ~ , \nonumber\\
    \Gamma_{\rm power}(T, q) & = & \Gamma(T, q) - \Gamma_0(T, q) \nonumber \\
    & = & \left( \frac{m}{2\pi T} \right)^{3/2} \left[ -\frac{1}{4} \omega^2T^2 + 
    \frac{q^2\omega^2}{24m}T^3 + \cdots \right] ~ . \nonumber
 \end{eqnarray}
In this work we will take into account all the terms in the square brackets for 
both $\Pi_{\rm power}(T)$ and $\Gamma_{\rm power}(T, q)$. 
Notice that each term is a power in $T$ and/or $q^2$.
Thus, retaining a fixed number of power corrections restricts the convergence of 
$\Pi_{\rm power}(T)$ and $\Gamma_{\rm power}(T, q)$ to the region of not too large 
values of $T$ and/or $q^2$, as it happens in QCD when the OPE series is truncated.

\section{Ground-state parameters}

Making use of the quark-hadron duality hypothesis, which assumes that the excited-state 
contribution is dual to the high-energy region of the free-quark diagrams, one gets the 
sum rules for $R_{\rm g}$ 
 \begin{eqnarray}
    \label{sr_2pt}
    R_{\rm g} e^{-{E_{\rm g}}T} = \Pi_{\rm power}(T) + \int\limits_{0}^{z_{\rm eff}^{\Pi}(T)} 
    dz ~ \rho_0(z) ~ e^{- z T} ~ , 
 \end{eqnarray}
and for the form factor $F_{\rm g}(q)$
 \begin{eqnarray}
    \label{sr_3pt}
    && R_{\rm g} F_{\rm g}(q) e^{-{E_{\rm g}}T} = \Gamma_{\rm power}(T, q) \\
    && \quad + \int\limits_{0}^{z_{\rm eff}(T, q)} dz_1 \int\limits_{0}^{z_{\rm eff}(T, q)} dz_2 ~ 
    e^{-\frac12(z_1 + z_2) T} \Delta_0(z_1, z_2 ,q) ~ , \nonumber
 \end{eqnarray}
where $\rho_0(z)$ and $\Delta_0(z_1,z_2,q)$ are the known spectral densities of 
the two- and three-point Feynman diagrams of the non-relativistic field theory 
\cite{lms_2ptsr,lms_3ptsr}. 

\begin{figure}[!htb]

\includegraphics[width=8cm]{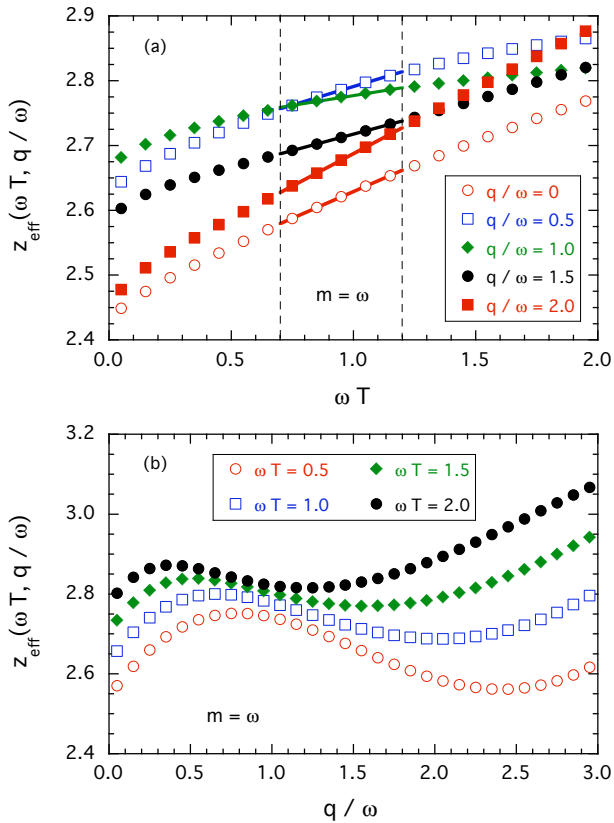}

\caption{\label{fig:zeff}
Effective continuum threshold $z_{\rm eff}(T, q)$ for the 3-point function, obtained 
by solving numerically Eq.~(\ref{sr_3pt}) using the exact bound-state parameters 
$R_{\rm g}$ and $F_{\rm g}(q)$  as well as the exact power expansion $\Gamma_{\rm power}(T, q)$, 
versus the Euclidean time $T$ at fixed values of the momentum transfer $q$ (a) 
and versus $q$ at fixed values of $T$ (b).
In (a) the vertical dashed lines identify the {\em fiducial range} in $T$ (see 
text), while the solid lines are linear fits of $z_{\rm eff}(T, q)$ in the 
fiducial range.}

\end{figure}

The relations (\ref{sr_2pt},\ref{sr_3pt}) constitute the definitions of the exact 
effective continuum thresholds $z_{\rm eff}^{\Pi}(T)$ and $z_{\rm eff}(T, q)$.
Their full $T$- and $q$-dependences can be obtained by solving Eqs.~(\ref{sr_2pt},\ref{sr_3pt}) 
using the exact bound-state parameters $R_{\rm g}$ and $F_{\rm g}(q)$ as well as 
the exact power expansions $\Pi_{\rm power}(T)$ and $\Gamma_{\rm power}(T, q)$. 
In the HO model this can be easily done numerically. 
Without loss of generality we set $m = \omega$ and show the corresponding results 
in Fig.~\ref{fig:zeff}.
It can clearly be seen that the effective continuum threshold $z_{\rm eff}(T, q)$ 
does depend upon both $T$~and~$q$.

Let us consider a restricted problem when the energy $E_{\rm g}$ of the ground state is 
known, and try to determine its elastic form factor from the sum rule (\ref{sr_3pt}).

First, according to \cite{svz} we should determine the Borel window (or the {\em 
fiducial range}), where the sum rule may be used for the extraction of the 
ground-state parameter: 
i) the lower boundary of the $T$-window is found from the requirement that the 
ground state gives a sizable (we require more than 50\%) contribution to the 
correlator; and
ii) the upper boundary of the $T$-window is obtained from the condition that the 
truncated OPE gives a good approximation to the exact correlator. 
Since in the HO model the power corrections are exactly known, the upper boundary 
is $T = \infty$. 
However, to be close to realistic situations when only a limited number of power 
corrections is available, we take from our study of Ref.~\cite{lms_3ptsr} the 
{\em fiducial range} $0.7 \lesssim \omega T \lesssim 1.2$ (see 
Fig.~\ref{fig:zeff}(a)) \footnote{We have checked that our findings are not sensitive 
to the specific choice made for the Borel window.}. 

Second, we must choose a criterion to approximate the effective continuum threshold 
$z_{\rm eff}(T, q)$. 
In this work we compare three different approximations:
  \begin{eqnarray}
      \label{constant}
      z_{\rm eff}(T, q) & \approx & z_0^C(q) ~ , \\
      \label{linear}
      z_{\rm eff}(T, q) & \approx & z_0^L(q) + z_1^L(q) ~ \omega T ~ , \\
      \label{quadratic}
      z_{\rm eff}(T, q) & \approx & z_0^Q(q) + z_1^Q(q) ~ \omega T + z_2^Q(q) ~ \omega^2 T^2 ~ .
  \end{eqnarray}
The standard procedure adopted in the sum-rule method is to assume a $T$-independent 
value, i.e.~Eq.~(\ref{constant}).

\begin{figure}[!htb]
\includegraphics[width=8cm]{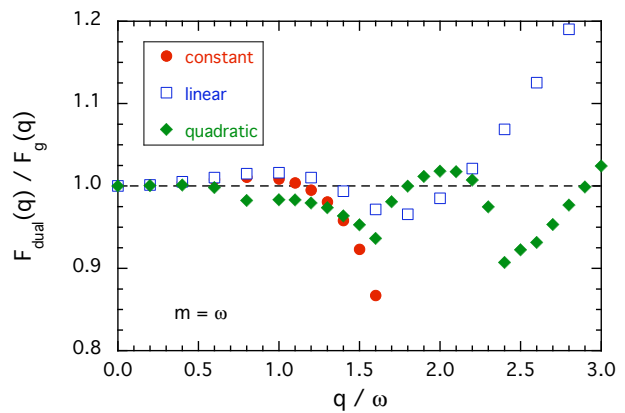}
\caption{\label{fig:FF}
Ratio of the dual form factor $F_{\rm dual}(q)$, extracted from the sum rule (\ref{sr_3pt}) 
using different approximations for $z_{\rm eff}(T, q)$, and the exact ground-state 
form factor $F_{\rm g}(q)$, given by Eq.~(\ref{FG}).
The dots, squares and diamonds correspond, respectively, to the results obtained using 
the constant (\ref{constant}), the linear (\ref{linear}) and the quadratic (\ref{quadratic}) 
approximations for the $T$-dependence of $z_{\rm eff}(T, q)$.}
\end{figure}

At each value of $q$ we fix the parameters appearing on the r.h.s~of 
Eqs.~(\ref{constant}--\ref{quadratic}) in the following way:
we define the dual energy, $E_{\rm dual}(T, q)$, as
 \begin{eqnarray}
     \label{Edual}
     E_{\rm dual}(T, q) =  - \frac{d}{d T} ~ \mbox{log} ~ \Gamma_{\rm dual}(T, q, z_{\rm eff}(T, q)) ~ ,
 \end{eqnarray}
where $\Gamma_{\rm dual}$ is the r.h.s.~of Eq.~(\ref{sr_3pt}) calculated using the 
approximations (\ref{constant}--\ref{quadratic}) for $z_{\rm eff}(T, q)$.
Then we calculate $E_{\rm dual}(T, q)$ at several values of $T = T_i$ ($i = 1,\dots, N$) 
chosen uniformly in the fiducial range and finally we minimize the squared 
difference with the exact value $E_{\rm g}$:
 \begin{eqnarray}
     \label{chisq}
     \chi^2 \equiv \frac{1}{N} \sum_{i = 1}^{N} \left[ E_{\rm dual}(T_i, q) - E_{\rm g}\right]^2 ~ .
 \end{eqnarray}
The results for the {\em dual} form factor $F_{\rm dual}(q)$, obtained via Eqs.~(\ref{sr_2pt}) and (\ref{sr_3pt}) 
using the correlator $\Gamma_{\rm dual}$ after optimizing the parameters of the three 
approximations (\ref{constant}--\ref{quadratic}), are shown in 
Fig.~\ref{fig:FF}.
Note that because of current conservation the form factor should obey the absolute 
normalization $F_{\rm dual}(q = 0) = 1$. 
We therefore require $z^{\Pi}_{\rm eff}(T) = z_{\rm eff}(T, q = 0)$, so that the 
r.h.s.~of Eqs.~(\ref{sr_2pt},\ref{sr_3pt}) coincide and the dual form factor is 
properly normalized. 

Let us consider first the case of the $T$-independent approximation (\ref{constant}).
The criterion of minimizing the $\chi^2$ (\ref{chisq}) leads to $z_0^C(q)$ 
for which the dual energy $E_{\rm dual}(T, q)$ differs from $E_{\rm g}$ by less than 
$0.1 \%$ and to the dual form factor $F_{\rm dual}(q)$ which is largely $T$-independent 
in the whole fiducial range.
Such a stability, usually referred to as the Borel stability, is often (erroneously)  
claimed to be the way to control the accuracy of the extracted form factor.
From Fig.~\ref{fig:FF} it can be seen that the $T$-independent approximation (\ref{constant}) 
works well (better than $2 \%$) at low values of $q$.
However, for $q \gtrsim 1.7\, \omega$ the T-independent ansatz does not work at all, since 
it is impossible to reproduce the ground-state energy in the fiducial range.

When the linear approximation (\ref{linear}) for $z_{\rm eff}(T, q)$ is considered, 
the form factor can be extracted also for $q \gtrsim 1.7\, \omega$. 
Note that the {\em exact} effective continuum threshold can be very well approximated 
by a linear function of $T$ in the whole fiducial range [see Fig.~\ref{fig:zeff}(a)]. 
Nevertheless, this is not a guarantee that one can extract the {\em exact} form factor:
deviations of the order of several percent can be produced after minimization of 
Eq.~(\ref{chisq}) up to $q \approx 2 \omega$ and uncertainties of the order 
of $10 \div 20 \%$ may plague the extracted form factor at $q \gtrsim 2 \omega$. 

One may try to go further and consider the quadratic Ansatz for $z_{\rm eff}(T, q)$ (\ref{quadratic}). 
However, as can be seen from Fig.~\ref{fig:FF}, this leads to certain instabilities in the extracted value of the form factor. 
These instabilities just reflect the fact that the unique solution to the problem of extracting 
the form factor from the correlator in a limited $T$-window does not exist \cite{lms_2ptsr}.  
Therefore, there is no way to get a systematic improvement in the accuracy of the extracted form factor by increasing 
the degree of the polynomial Ansatz for $z_{\rm eff}(T, q)$.

Nevertheless, it is worth emphasizing that in the HO model the comparison between the form factor 
extracted assuming Eq.~(\ref{linear}) and the one obtained using Eq.~(\ref{quadratic}) 
gives a realistic estimate of the accuracy in a wide range of values of $q$.
Whether this feature persists in QCD is an interesting and important issue to be addressed 
in the future.
\section{Conclusions}
Let us summarize the main messages of our analysis:
%\begin{itemize}
%\item 
$\bullet$ 
The knowledge of the correlator in a limited range of relatively small Euclidean 
times $T$ (that is, large Borel masses) is not sufficient for the determination of the
ground-state parameters. 
In addition to the OPE for the relevant correlator, one needs an independent criterion 
for fixing the effective continuum threshold. 
%\item 

$\bullet$
Assuming a $T$-independent (i.e., a Borel-parameter independent) effective 
continuum threshold the error of the extracted ground-state parameter (both decay 
constant and form factor) turns out to be typically much larger than 
(i) the error of the description of the exact correlator by the truncated OPE and 
(ii) the variation of the bound-state parameter in the fiducial range (i.e., Borel window).
The latter point is of particular relevance since the Borel stability is usually 
believed to control the accuracy of the extracted ground-state parameter. 
Obviously, this is not the case (see also Refs.~\cite{lms_2ptsr,lms_3ptsr,m_lcsr}).
%\item 

$\bullet$
Allowing for a $T$-dependent effective continuum threshold and fixing it according to Eq.~(\ref{chisq}) 
leads to evident improvements in the extracted ground-state parameters. 
This was shown for the decay constant and the form factor in the HO model. 
Moreover, in this model the variation of the form 
factor extracted using different approximations for $z_{\rm eff}$ gives {\it de facto} 
a realistic error estimate. Unfortunately, the use of higher polynomial approximations leads to 
instabilities in the fitting procedures. It is, therefore, impossible to construct a systematic 
procedure which would converge to the exact effective continuum threshold. 
As the result, rigorous error estimates cannot be obtained. 

%\end{itemize}

The impossibility to get a rigorous control over the systematic errors of the extracted 
ground-state parameters is the weak feature of the method of sum rules and an obstacle 
for using the results from sum rules in problems where rigorous error estimates are required. 

In spite of this weakness, the application of the proposed modifications of the method 
in QCD seems very promising and may lead to a considerable increase of the {\it actual} 
accuracy of the calculated hadron parameters. 
This issue deserves a serious investigation. 

%
%Finally, we want to comment on the obtained quantitative estimates. 
%In the HO model the ground state is well separated from the first excitation that  
%contributes to the correlator by a large energy gap equal to $2 \omega$. 
%This feature makes the HO model a very favorable case for the application of sum rules. 
%Whether or not a comparable accuracy may be achieved in QCD, where the corresponding 
%feature is absent, is an open question.
%

\vspace{.2cm}
{\it Acknowledgments:} D.~M.~was supported by the Austrian Science Fund (FWF) under project P20573 
and grant of the President of Russian Federation 1456.2008.2.
% for leading scientific schools.


\begin{thebibliography}{30}
\bibitem{footnote}
Ref.~\cite{svz} with about 3500 citations is among the top cited papers in high energy physics.
\end{thebibliography}
\end{document}